# SENTIMENT ANALYSIS BY USING FUZZY LOGIC

Md. Ansarul Haque[1], Tamjid Rahman [2]

[1]Department of Computer Science and Engineering, Stamford University, Bangladesh
[2]Department of Computer Science and Engineering, Stamford University, Bangladesh

## ABSTRACT

*How could a product or service is reasonably evaluated by anyone in the shortest time? A million dollar question but it is having a simple answer: Sentiment analysis. Sentiment analysis is consumers review on products and services which helps both the producers and consumers (stakeholders) to take effective and efficient decision within a shortest period of time. Producers can have better knowledge of their products and services through the sentiment analysis (ex. positive and negative comments or consumers likes and dislikes) which will help them to know their products status (ex. product limitations or market status). Consumers can have better knowledge of their interested products and services through the sentiment analysis (ex. positive and negative comments or consumers likes and dislikes) which will help them to know their deserving products status (ex. product limitations or market status). For more specification of the sentiment values, fuzzy logic could be introduced. Therefore, sentiment analysis with the help of fuzzy logic (deals with reasoning and gives closer views to the exact sentiment values) will help the producers or consumers or any interested person for taking the effective decision according to their product or service interest.*

## 1. INTRODUCTION

Now-a-days, time and reliable source is very much needed to gather the deserving information related to any specific matter. Web in one sense can provide those deserving information maintaining the less time and reliable source. Opinion is the vital type of information on the web. These opinions are expressed in some user generated contents such as customer reviews of products, micro-blogs, and forum posts. So, this is referred as online 'word-of-mouth'.

Social media refers to the web-based technologies which turns the communication into an interactive dialogue. These media are usually used for social interaction. These provide a huge information about different individual's interest and behaviors and also retrieve all the information related to certain events. After retrieving, we can distinguish what is important and what is negligible. [1] [2]

Among the top-ranked social networking sites, twitter (which launched in 2006) is very popular for its micro-blogging features. Its information helps to answer the technological and sociological queries. In this modern age, it is too expensive and time consuming to proceed without this type of social network. Within a short period, around 160 million users are cope up with its service, specifically saying with its allocation of 140 characters. Just for an example we could refer that during the period of earthquake in Indonesia Twitter has given its feedback and played the key role with the performance which was greater than the other electronic media such as television, newspapers and so on. Its vast flow of information helps to measure and analyze the users' opinions regarding technological, social, environmental and other issues. [3]

## 1.1 Sentiment Analysis

Everyday millions of comments or opinions are posted in websites that provide the facilities for micro-blogging such as Twitter or Facebook. The creators of the comments share their opinions on different topics, discuss current issues even spot accidents or any flu outbreaks. These are the valuable source of opinions and sentiments as huge amount of posts are posted by the users according to their used products and services, or express their different views on different perspectives. Researchers are using these posts to measure the public sentiment and to do sentiment analysis. They are trying to determine the "PN-polarity" of subjective terms i.e, identifies whether a term expresses the opinion which could have positive or negative connotation.

The purpose of sentiment analysis is based on the two sectors:

i. **Classifying Documents**

Classifying documents or any passages according to sentiment orientation such as positive vs. negative.

ii. **Gathering Information**

Extracting information of opinions which contains information of particular aspects of interest and the corresponding sentiment orientation in a structured form from a set of unstructured data.

The tasks of classifying documents of the sentiment analysis can be divided into three sub-tasks:

i. Identifying SO polarity:

Whether the comment or post is referring a situation or event without disclosing the subjectivity (positive or negative opinion) on it or expressing opinion on its subject matter. Briefly, it means that identify the subjective or objective polarity of a post or comment.

ii. Identifying PN-polarity:

Whether a subjective post or comment is expressing positive or negative.

iii. Identifying the Degree of PN-polarity:

This step gives the impression of the degree of positivity or negativity on that opinion. Positivity could be weakly positive, mildly positive or strongly positive and same could be for the negative opinion.

"I think this is going to be one of the most important datasets of this era, because we are looking at what people are talking about in real time at the scale of an entire society," says Mislove, an assistant professor of computer science at Northeastern University. So, sentiment analysis is doing the right thing as he told. [13][14] [2].

## 1.2 SentiWordNet

SentiWordNet is the lexical (converts a sequence of characters into a sequence of tokens) resource for sentiment analysis in which three numerical scores are maintained by Pos(), Neg() & Obj() which represents how much positivity, negativity & objectivity are contained in those opinions. This is called sentiment classification which determines the subjectivity of a given text. It is the process of deciding whether a given text expresses a positive or negative opinion about its "subject matter" and "subject attributes", which also known as 'product' and 'features'. It focuses on the quantitative analysis. It is very much popular and free for the research works. It reflects a nice outlook in its graphical user interface.

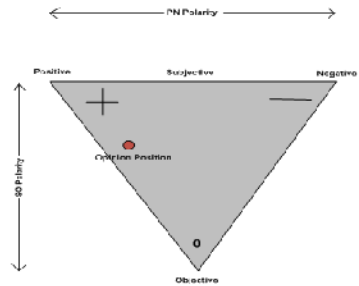

Fig 1.1: Graphical Representation Adopted by SentiWordNet

Considering the SentiWordNet (version 1.0), the synset are containing three numerical scores Pos(s), Neg(s) & Obj(s). These numerical scores have the range from 0 to 1 and the sum of all these scores is 1 for each synset. [14].

## 1.3 Twitter as a Micro-blogging Website

Micro-blogging is very popular communication tool among the internet users and one kind of information center. In this micro-blog, people posts their real-time comments about their opinions on a variety of topics, complain, ideas, discuss different issues and feel free to express their sentiments regarding any products and services. Micro-blogging is getting popular and replacing the traditional blogs as traditional blogs or mailing lists are not providing the free format of messages & easy accessibility as the micro-blogging platforms are used to provide. For this reason, twitter as a micro-blogging website, it is sometimes called as "The SMS of the internet".

At present twitter is very popular micro-blogging websites considering the duration, impression and popularity acquired by itself. It has launched on July, 2006. It's a social networking service that allows users to post real time messages and their opinions, called tweets. It is having restriction of 140 characters in length and having no headache of misspellings, slangs or abbreviations. It mainly focuses on individuation and characterization of opinions in a text. For this reason, it is very much efficient in the field of sentiment analysis or opinion mining. It contains a very large number of very short messages and the contents vary from the personal thoughts to public expression. It classifies the tweets into three sentiments: positive, negative and neutral. [11] [12].

The terminology is associated with tweets are as follows:

i. **Emoticons:** Facial expressions are the pictorial representations which express the user's mood. It also contains the punctuations and letters. Ex: ☐, :D, C:, ☐, D8, D; , : | where first

three are representing positivity, next three are representing negativity and last two are representing neutrality.
ii. **Target:** Users can use '@' symbol to refer any other twitter user on the micro-blog and Twitter automatically alerts them. Ex: I love to thank you for proposing that iPhone@John

| Abbreviations | Meaning |
|---|---|
| Lol | laughing out loud |
| gr8 | Great |
| Bff | best friend forever |
| Rotf | rolling on the floor |

iii. **Hash tags:** Users can use hash tags to mark topics to increase the visibility of their tweets. Ex: Results for #electronic-media. [11].

Mostly used abbreviations or short-terms in twitter are:

## 1.4 Twitter API

API stands for Application Programming Interface and it is a defined way for a program to establish a task especially by retrieving or modifying data. Twitter API helps the programmer to make projects, applications or websites that interact with Twitter. As a user we use browsers through HTTP (Hyper Text Transfer Protocol) to visit and interact with the websites where programs talk to the Twitter API over HTTP. In our project we have used Twitter 4J as for the Twitter API which is the unofficial Java library.

Accessing twitter through API is ten times more efficient than the web interface. For this purpose, different users can use different API, such as:

i. **Desktop users**: twitterrific & twhirl
ii. **Cell users**: TinyTwitter, PocketTweets & iTweets   [15]
   have accessed these databases through the programming commands.

## 2. Proposed Techniques of Sentiment Analysis

### 2.1 Introducing Required Tools

**i. SentiWordNet**

In our project we have used the **SentiWordNet 3.0.0** for having the values of the user's opinions or sentiments. Basically it is using the positive and negative values. For the zeros of both (positive and negative), we are considering the neutral opinion. We have described about the SentiWordNet in the earlier sections (section 3.2).

**ii. Twitter Website**

From the twitter (www.twitter.com) we have extracted the tweets to analyze them. As for example, in twitter we can search for any specific topic or product such as iPhone. Different individuals have different thoughts of their products and they share their positive and negative opinions. These tweets are useful for the analysis and for our project.

**iii. Java (Eclipse 3.7.1)**

For developing applications in Java, Eclipse is the well-suited software development environment. It comprises with Integrated Development Environment (IDE) and various plug-ins. We have chosen to work with the Eclipse 3.7.1 among the different versions.

**iv. Databases**

We are using the databases in storing tweets after extracting and also the SentiWordNet words with the values. Databases are in the form of Microsoft Excel Files (such as .xls or .txt).

## Our Project Procedure

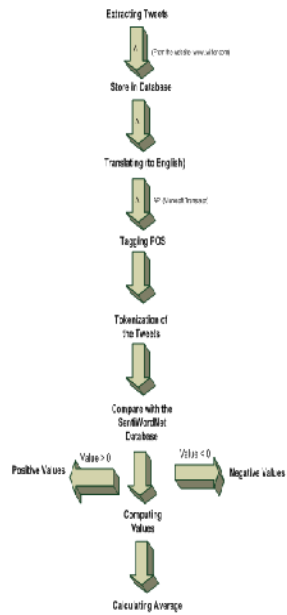

Fig 2.1: Sentiment Analysis Procedure

**2.2.1 Searching Tweets**

We have used the website www.twitter.com to have the frequent tweets from the users. As per we know that tweets are the comments or posting of the users in the Twitter. The users can post their tweets depend on their ideas, concepts or their likings about any product or service. We have used Twitter 4j for retrieving tweets from the Twitter website and at a time we are extracting 100 tweets by using this package.

As our concentrating item of the project is "iPhone", we went through that term. Whatever we get, we can classify them in two forms

   a. Subjective and
   b. Objective

Subjective means it is representation of the mixture of positive and negative values or sets and on the other hand, the objective means representation of neutral values (0) or sets. In our project, we

37

have skipped the neutral values or sets. Later we used the POS-tagging where we used the parts-of-speech for tagging the tokens. Subjective texts tend to use base form of verbs (VB) and also simple past tense (VBD) instead of the past participle (VBN). Adverb (RB) is mostly used in subjective texts to give an emotional color to a verb.

### 2.2.2 Saving the Extracted Tweets

After having the tweets, we have saved those tweets in text files (such as "TempTweets.txt") by using FileWriter application. At present, we have so many assisting tools to extract tweets from different sources. As for example, through programming we could use API which is facilitated by popular environment such as Java or by using well-established software such as Archivist.

### 2.2.3 Reading the File (where Tweets are stored)

FileInputStream fstream = new FileInputStream("C:/Users/stalin/Desktop/Temptweets.txt");

By using the FileInputStream, we accessed the Temptweets.txt which one was created after extracting and saving the extracted tweets. Here we constructed the object of the FileInputStream which extract data from the excel file. Then we used BufferReader for reading line by line.

### 2.2.4 Translating the Tweets to English

Translate.setKey("2768f0575d056bb86c91a4b0cf588e1d7382c15a");String translatedText = Translate.execute(strLine,Language.ENGLISH);

The tweets could be from different users of different nationalities in different languages. Those tweets of different languages except English are also important enough to analyze sentiments. There are so many translators with API to translated in English language in our real life. For our project, we have used Microsoft Translator to get those tweets in English and we have used the API key .

("2768f0575d056bb86c91a4b0cf588e1d7382c15a") to translate tweets.

### 2.2.5 Tagging the Tweets

MaxentTagger tagger = new MaxentTagger("D:/java prog/FLproject/taggers/bidirectional-distsim-wsj-0-18.tagger");

String tagged = tagger.tagString(translatedText);

For the sake of our sentiment analysis, at first we tagged the tweets according to the parts-of-speech such as noun, verb, adjective, etc. As in our database (SentiWordNet), we have different focus on different parts-of-speech. As prepositions and conjunctions are too common to be used among statements, we can remove them easily after tagging as well as proper nouns which usually don't have an affective content. After removing the unaffecting content, we usually have four POS: adjective, verb, adverbs and noun which are known as opinion words.

POS-tagger is an application that helps to read text in some languages and assign part-of-speech to individual terms, words or tokens. The tagger what we have used here is written by The Stanford Natural Language Processing Group in Java. Among its three models, we have used the English tagger model. POS tagger which we used is named as "Penn Treebank Tagset":

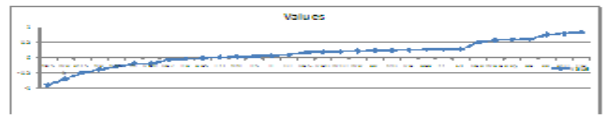

Fig 2.2. Values for Objective vs. Subjective

Here, WP$ → Possessive wh-pronoun
POS → Possessive Ending
NNS → Noun Plural
IN → Preposition
VBN → Verb, past participle
VBZ → Verb, $3^{rd}$ person singular present
JJR → Adjective, comparative
MD → Modal
RBS → Adverb, superlative
TO → to
NN → Noun, singular or mass
JJS → Adjective, superlative
JJ → Adjective
DT → Determiner
VBG → Verb, gerund
VBD → Verb, past tense
WDT → Wh-determiner
WP → Wh-pronoun
RP → Participle
VB → Verb, base form RBR
RBR → Adverb, comparative
CC → Coordinating Conjunction
Ex → Extential there
VBP → Verb, non-$3^{rd}$ person
WRB → Wh-adverb
RB → Adverb
PDT → Pre-determiner
UH → Interjection

We have given some cases for the above figure:

i. Objective texts contain more common and proper nouns.
ii. Verbs in objective texts are usually in the third person and used more often in past participle.
iii. Superlative adjectives are used more for expressing emotions and opinions, comparative adjective are used for starting facts.
iv. Verbs in base form are used with modal verbs to express emotions.
v. Authors of subjective texts usually write about themselves (verbs in first person)or address the audience (second person) and tend to use simple past tense.
vi. Subjective texts contain more personal pronouns.
vii. Utterances are strong indicators of a subjective text.

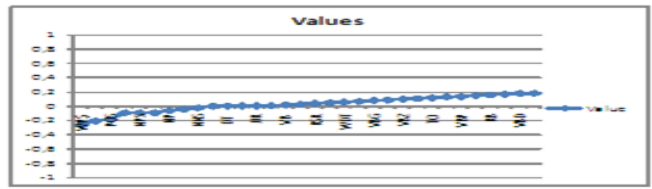

Fig 2.3: Values for Positive vs. Negative

### 2.2.6 Removing Punctuation Marks

String pure = translatedText.replaceAll("\\w+@\\w+\\.[\\w]{3} ","");
String strippedInput = pure.replaceAll("[!@#(:]","");

As the punctuation mark will not have any effect on the sentiment analysis, we could remove those marks whatever belongs to the tweets. It can also be called as filtering. Here in our project we tried to apply some punctuation marks such as @, #, !, ',', (: and so on. These removing of the punctuation marks would help to tokenize our tweets perfectly and efficiently. Such example of tweets:

>iPhone made me friendly @david

Here, we would like to remove the twitter user name (david) which is having the symbol @ giving nothing about opinions only tagged that named person.

Hashtags are words or phrases used by users to group posts together by topic or type, and they are prefixed by "#" as in example #iphone4s. As they are not used to express opinions, we can remove them before the analysis.

### 2.2.7 Tokenization

String[] a=strippedInput.split(" ");

According to our project, tokens are the meaningful terms or elements of text. Tokens can be any words, symbols or phrases. Tokenization is the process of splitting a stream of text up into those forms of tokens. This process is very much related to sentiment analysis as our database which is referring as SentiWordNet works with those tokens. For our simplification, we have changed all the tokens into lower-case.

### 2.2.8 Removing Stop Words

```
if(aList1.contains(stopwordArr));
for(int i=0;i<stopwordArr.length;i++)
{
aList1.remove(stopwordArr[i]);
}
```
Stop words are those which are non-relevant of our sentiment values. If any tweet is having those stop words, it will be removed from the list. This is applied for all the extracting tweets. The list words of stop words we have used in our project are so many. Among them we would like to pick-up some of them: a, about, above, after, again, against, all, am, an, and, are, aren't, at, be, because, been, before, being , below, between, both, but, by, can't, cant, could, couldn't, do, does,

40

doesn't, did, didn't, doing, down, during, each, below, for, further, has, hasn't, have, haven't, he'll, he's, her, here, here's, hers, herself, himself, him, himself, it's, its, let's, me, more, mustn't, my, myself, no, nor, off, on, of, what, what's, when, where, where's, which, who's, why, whom, why's, with and so on

### 2.2.9 Creating Database Connection with the File

Class.forName("sun.jdbc.odbc.JdbcOdbcDriver");
Connection con = DriverManager.getConnection("jdbc:odbc:TempData");
Statement stmt = con.createStatement();

We have created connection with our database (TempData which is having SentiWords with positive and negative values) which works as SentiWordNet.

### 2.2.10 Storing Splitting Tweets in Array

String[] s1 = aList1.toArray(new String[aList1.size()]);
All the splitting tweets should be listed in array for simplification of calculating scores.

### 2.2.11 SQL for Extracting Scores

SQL = "SELECT PosScore,NegScore FROM Sheet1 WHERE SynsetTerms='" + s1[j] + "'"  ;
stmt.execute(SQL);
 ResultSet rs = stmt.getResultSet();

  The database has two columns for the positive and negative scores of the sentiment terms. Through the SQL (Structured Query Language), we accessed the positive and negative scores of the each term of the tweets.

### 2.2.12  Calculating Total Scores

Score = (Positivescore - Negativescore);

After getting positive and negative scores of the each term of the tweet, we calculated the total positive score and total negative scores. Then we deducted each other and got the final score.

### 2.2.13  Normalization

Normalization helps to organize data and it also minimizes redundancy. It produces well-structured relations in the database.

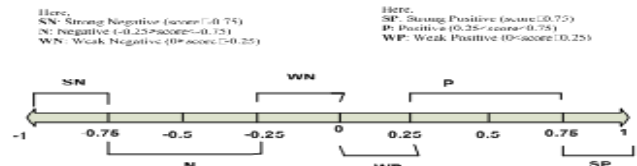

Fig 2.4: Normalization

### 2.2.14 Implementing Weights

Weights have given to the frequently used terms in our project. As we have concentrated on 'iPhone', we have given such weights to the mostly used terms. For example, 0.95 for iPhone, 0.9 for iPhone4s, 0.85 for iPhone4g.

### 2.2.15 Identifying and Categorizing the Positive and Negative Tweets

If the total score (positive score-negative score) is greater than 0, we will consider it as a positive tweet and if it is smaller than 0, we will consider it as an negative tweet. In this approach, we will count the number of positive and negative tweets. But the positivity and negativity depends on their tweet scores.

For the further classification which depends on tweet score, we have six different classes of tweets.

   a. Strong Positive Tweets
   b. Positive Tweets
   c. Weak Positive Tweets
   d. Weak Negative Tweets
   e. Negative Tweets
   f. Strong Negative Tweets

### 2.2.16 Computing the Results

In our project, we tried to analyze sentiments on the tweets of 'iPhone' and at last we calculated the following results:

 **a. No. of Tweets**
Counting the number of tweets to be processed.

 **b. Total no. of Positive Tweets**
Let, t is the token or words and sent(t) is the sentiment value. Then we can take the tweet as the positive tweet if

$\{(t \in Tweets) \;\&\&\; (sent(t) > 0)\}$

 **c. Total no. of Negative Tweets**
Let, t is the token or words and sent(t) is the sentiment value. Then we can take the tweet as the negative tweet if

$\{(t \in Tweets) \;\&\&\; (sent(t) < 0)\}$

 **d. Weighted Mean**
Weighted Mean = $(\sum_{i=0}^{n} w_i a_i) / (\sum_{i=0}^{n} w_i)$
Here, $w_i$ = weight and $a_i$ = value of the tweet

If we want to give more important values to the main features of any product or service, we could use weighted mean. Descriptive statistics uses this mean very precisely. It is given priority more focus to the main characteristics more than other usual characteristics.

### e. Arithmetic Mean

Arithmetic Mean = (Total Values of the Tweets / Total No. of Tweets)
Arithmetic mean is giving equal importance to all data and this is the basic difference between weighted and arithmetic mean.

### f. Positive Sentiment by Percentage

PS (%) = (Number of Positive Sentiments/Total Number of Tweets) *100

### g. Negative Sentiment by Percentage

NS (%) = (Number of Negative Sentiments/Total Number of Tweets) *100

**Our analysis could give such results as in the following diagram:**

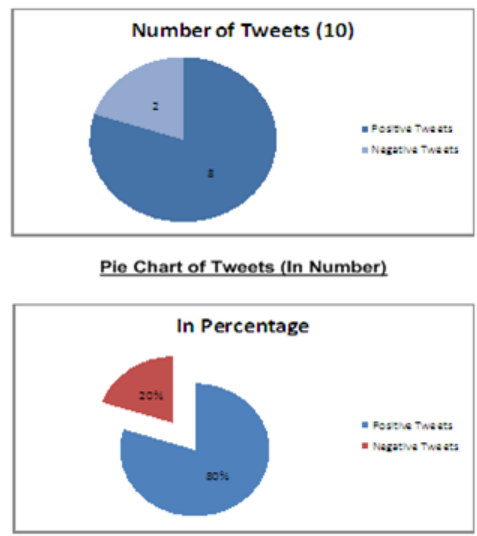

Fig 2.5: Pie Chart of Tweets (In Percentage)

## 3. RESULT DISCUSSION

Here for the testing purpose of our project, we have extracted six (10) tweets regarding ''iPhone'' and they are displaying in our output screen in the following form:

### Input:

```
@jasonenriquez:J'aime mon Iphone4S.
@stalin:iphone 4s is lovely!!
@YasminScott98:Touch-screen of iphone@ is lovely http://t.co/HY1zqtzq
and attractive
@JessMarieFrench:naked iphone(: is catchy and shiny
@dgrey1986:iphone4s is sloppy in battery
@hlouisewagg:Damn iphone
@BxDiimegambler:iphone is Not bad
```

```
@SabrinaHu5:iphone is not good
@nash711:nokia 4 is good
@GersForum:So I just got the iPhone 4s and it's amazing :)
```

Here, we have extracted 10 tweets as sample. The tweets are initialized with the username followed by their given comments. From the first tweet, we can see its in French and others are in English. We have shown our output below and we can find out how our project is working for different languages, its tagging, calculating the tweet values and also showing the tweets status according to their degree of positivity and negativity.

**Output:**

```
Loading default properties from trained tagger D:/java
prog/FLproject/taggers/bidirectional-distsim-wsj-0-18.tagger
Reading POS tagger model from D:/java
prog/FLproject/taggers/bidirectional-distsim-wsj-0-18.tagger ... done
[12.9 sec].
```

**@/SYM jasonenriquez/NN :/: I/PRP love/VBP my/PRP$ Iphone4S/NNS ./.**

0.25     positive

```
Loading default properties from trained tagger D:/java
prog/FLproject/taggers/bidirectional-distsim-wsj-0-18.tagger
Reading POS tagger model from D:/java
prog/FLproject/taggers/bidirectional-distsim-wsj-0-18.tagger ... done
[3.2 sec].
```

**@/IN stalin/NN :/: iphone/NN 4s/NNS is/VBZ lovely/JJ !!/NN**

0.25     positive

```
Loading default properties from trained tagger D:/java
prog/FLproject/taggers/bidirectional-distsim-wsj-0-18.tagger
Reading POS tagger model from D:/java
prog/FLproject/taggers/bidirectional-distsim-wsj-0-18.tagger ... done
[2.5 sec].
```

**@/IN YasminScott98/NNP :/: Touch-screen/NN of/IN iphone/NN @/SYM is/VBZ lovely/JJ http://t.co/HY1zqtzq/JJ and/CC attractive/JJ**

0.375     positive

```
Loading default properties from trained tagger D:/java
prog/FLproject/taggers/bidirectional-distsim-wsj-0-18.tagger
Reading POS tagger model from D:/java
prog/FLproject/taggers/bidirectional-distsim-wsj-0-18.tagger ... done
[3.3 sec].
```

**@/IN JessMarieFrench/NNP :/: naked/JJ iphone/NN -LRB-/-LRB- :/: is/VBZ catchy/JJ and/CC shiny/JJ**

0.375     positive

```
Loading default properties from trained tagger D:/java
prog/FLproject/taggers/bidirectional-distsim-wsj-0-18.tagger
```

```
Reading POS tagger model from D:/java
prog/FLproject/taggers/bidirectional-distsim-wsj-0-18.tagger ... done
[3.4 sec].
```

**@/IN dgrey1986/CD :/: iphone4s/NNS is/VBZ sloppy/JJ in/IN battery/NN**

```
0.1875    weak_positive
Loading default properties from trained tagger D:/java
prog/FLproject/taggers/bidirectional-distsim-wsj-0-18.tagger
Reading POS tagger model from D:/java
prog/FLproject/taggers/bidirectional-distsim-wsj-0-18.tagger ... done
[3.4 sec].
```

**@/SYM hlouisewagg/NN :/: Damn/JJ iphone/NN**

```
-0.75    negative

Loading default properties from trained tagger D:/java
prog/FLproject/taggers/bidirectional-distsim-wsj-0-18.tagger
Reading POS tagger model from D:/java
prog/FLproject/taggers/bidirectional-distsim-wsj-0-18.tagger ... done
[3.1 sec].
```

**@/IN BxDiimegambler/NNP :/: iphone/NN is/VBZ Not/RB bad/JJ**

```
0.375    positive

Loading default properties from trained tagger D:/java
prog/FLproject/taggers/bidirectional-distsim-wsj-0-18.tagger
Reading POS tagger model from D:/java
prog/FLproject/taggers/bidirectional-distsim-wsj-0-18.tagger ... done
[3.3 sec].
```

**@/IN SabrinaHu5/NNP :/: iphone/NN is/VBZ not/RB good/JJ**
```
-1.0    negative

Loading default properties from trained tagger D:/java
prog/FLproject/taggers/bidirectional-distsim-wsj-0-18.tagger
Reading POS tagger model from D:/java
prog/FLproject/taggers/bidirectional-distsim-wsj-0-18.tagger ... done
[3.2 sec].
```

**@/IN nash711/CD :/: nokia/NN 4/CD is/VBZ good/JJ**

```
0.625    positive

Loading default properties from trained tagger D:/java
prog/FLproject/taggers/bidirectional-distsim-wsj-0-18.tagger
Reading POS tagger model from D:/java
prog/FLproject/taggers/bidirectional-distsim-wsj-0-18.tagger ... done
[3.3 sec].
```

**@/IN GersForum/NNP :/: So/IN I/PRP just/RB got/VBD the/DT iPhone/NNP 4s/NNS and/CC it/PRP 's/VBZ amazing/JJ :/: -RRB-/-RRB-**

```
0.6875    positive

Total no of tweets is:10.0
Total no of positive tweets:8.0
```

```
Total no of negative tweets:2.0
Arithmetic mean is:0.1375
Sentiment by Percent
Positive sentiment % is:80.0
Negative sentiment % is:20.0
```

First of all, the POS tagger is generated and starts tagging the extracted tweets. Then computing the sentiment values of the tweets are done and given their sentiment status. By having the values of the tweets and the weights, we can compute the weighted and arithmetic mean. Then we can have the percentage of the individual sentiments (positive and negative).

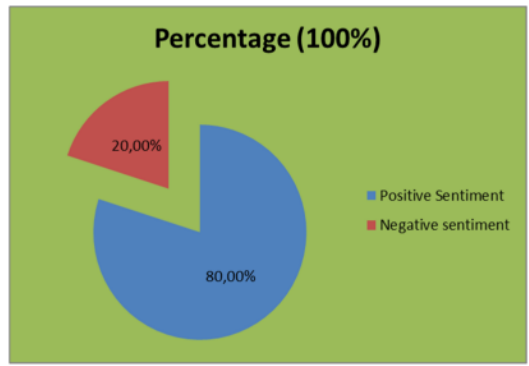

Fig 3: Percentage of the Sentiments

A big problem regarding the word-level analysis is that detecting the use of irony and sarcasm is very hard, as a whole understanding of the sentence is required for that. Still, we have tried to improve this word-level analysis and trying to detect negation particles before adjectives. For example, if we analyze the words "iphone is not bad", it will get an overall negative value for 'not' and 'bad'. But the combination of those words is supposed to have a positive meaning. So, if we can detect those negation particles, we can invert the value of the following adjective.

## 4. ANALYSIS

In this paper, we did analysis on 100 tweets. But as we could recommend that, the more the sentiments the more bold the analysis result. For our future works, we would like to extend our work with more tweets and have more robust result. This result would benefit the interested users with strong beliefs.

One of the limitations of our project is that it is focused on the sentiment classification and concentrating to manipulate the results according to those classifications (positive or negative). It is not focusing on the feature based classification. In our further steps, we will try to include this limitation and make stronger in this context.

We would like to concentrate on the analysis of context and domain, as both of them has the capability to influence the word or sentiment or opinion's attitude. As different individual could define their concepts or ideas in different manner, so it is really meaningful if we could cope up with that matter.

Sentiment analysis is really a challenging work for some critical perspective such as colloquial languages or opinions or for the ironic words from different reviewers. Our future works will have some steps on those obligations and try to get proper solutions.

For the ranking perspective, we could rank the reviewers for their most positive or negative posts or sentiments. We mean that have some feedbacks from the users normally and we would rank their feedbacks by manipulating the values of positivity or negativity of the sentiments. Then the interested entities could have the like or dislike portions on which they should concentrate hierarchically.

## 5. CONCLUSIONS

Sentiment analysis is nothing but special field of text analysis. In short, focus and analyze the extracted opinions (sentiments or emotional contents) from the posted comments. Our project goal is to analyze the sentiments on a topic which are extracted from the Twitter and conclude a remark (positive/negative) of the defined topics. We have implemented an easier procedure to analyze sentiments on any interested field or topic. Hope this project would helpful for anyone in any way to meet up their interests or what they deserve. This is our major goal of this project and waiting to provide much more worthy works in our future work.

## ACKNOWLEDGEMENT

We would have pleasure to thank the referees, who provided very useful and efficient resources. Without their insight resources and information, It would be very difficult to have our such position and project. Its really hard to express our gratitude level to them for their immense contributions.

Special thanks to David L. Hicks for his timely supervision and his benevolent guidelines. He was so caring and ready to help regarding the project, its objectives and aim, report writing and editorial guidance and so on.

We are also very much thankful to Henrik Legind Larsen. Very much effective part of our project is to have him and have proper direction in such manner to fulfill this project in the scheduled time with huge positive feedback.

No comments for the friends of other groups especially Damien, Emanuel, Sayeedi and many others.

At last we worked as a good communicating folk who interact and discuss each other to make this project properly. We invested our efforts and hard labor. We tried to make it more perfect and robust without any error or fault.

All of the remaining errors and faults are, of course, our own.
.

International Journal of Computer Science, Engineering and Information Technology (IJCSEIT), Vol. 4,No. 1, February 2014

## Authors

**Md. Ansarul Haque** is a faculty of Stamford University, Bangladesh. He received his M.Sc degree in IT, Halmstad University, Sweden and B.Sc in CSE from Shahjalal University, Bangladesh. His research interest lies in wireless communication, networking and fuzzy systems.

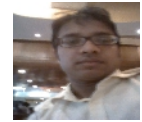

**Tamjid Rahman** is a faculty of Stamford University, Bangladesh and he received his BS degree in Computer Science and Engineering from Stamford University Bangladesh in 2010. He is a MS student of North South University. His research interest lies in artificial intelligence, software engineering, programming.

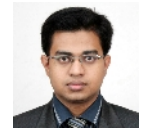